\documentclass[letter]{aa} 
\usepackage{graphicx}
\usepackage{txfonts}
\begin{document}
   \title{WSRT Faraday tomography of the Galactic ISM at
   $\lambda \sim 0.86$ m}
   \authorrunning{Schnitzeler et al.}

   \subtitle{First results for a field at ($l,b$)=(181\degr,20\degr)}

   \author{D.H.F.M. Schnitzeler\inst{1}
          \and
          P. Katgert\inst{1}
          \and
          A.G. de Bruyn\inst{2,3}
          }

   \offprints{D. Schnitzeler}

   \institute{Leiden Observatory, P.O. Box 9513, 2300 RA Leiden, The Netherlands\\
              \email{schnitze@strw.leidenuniv.nl}
         \and
             ASTRON, P.O. Box 2, 7990 AA Dwingeloo, The Netherlands
         \and 
             Kapteyn Institute, P.O. Box 800, 9700 AV Groningen, The Netherlands
             }

   \date{Received 13 April 2007 / Accepted 6 June 2007}

 
  \abstract
   {}
   {We investigate the distribution and properties of Faraday rotating and synchrotron emitting regions in the Galactic ISM in the direction of the Galactic anti-centre.
}
   {We apply Faraday tomography to a radio polarization dataset that we obtained with the WSRT.
We developed a new method to calculate a linear fit to periodic data, which we use to determine rotation measures from our polarization angle data. From simulations of a Faraday screen + noise we could determine how compatible the data are with Faraday screens.
}
   {An unexpectedly large fraction of 14$\%$ of the lines-of-sight in our dataset show an unresolved main component in the Faraday depth spectrum. For lines-of-sight with a single unresolved component we demonstrate that a Faraday screen in front of a synchrotron emitting region that contains a turbulent magnetic field component can explain the data.}
   {}

   \keywords{magnetic fields -- radio continuum:ISM -- ISM: magnetic fields, structure  -- techniques: polarimetric, miscellaneous
               }

   \maketitle
%

\section{Introduction}

Faraday tomography is a very powerful tool for studying the relative
line-of-sight distribution of regions with synchrotron emission and
Faraday rotation. If the thermal plasma does not emit its own
synchrotron radiation it simply acts as a Faraday rotating screen for
any linearly polarized radiation that illuminates it from the
back. However, if the thermal plasma co-exists with relativistic
particles which emit (linearly polarized) synchrotron radiation, the
situation can be much more complex. This is because the contributions
from various parts of the line-of-sight through the plasma combine
vectorially, which may even lead to complete cancellation (at certain
wavelengths) of the polarized signal emitted in the plasma.

For a simple Faraday screen, the amount of rotation of the plane of
linear polarization gives direct information on the Faraday depth
$\mathcal{R}$\ [rad/m$^2$] of the screen:
$\mathcal{R} = 0.81\int\limits_{source}^{observer}
n_e\vec{B}\cdot\mbox{d}\vec{l}$, the line-of-sight integral of the product of the 
electron density [cm$^{-3}$] and the magnetic field component [$\mu$G] along the
line-of-sight [pc]. In that case, $\mathcal{R}$ and the rotation measure
$RM=\partial \Phi/\partial \lambda^2$ (where $\Phi$ is the
polarization angle of the radiation) are identical. In the more
general case, with a mixture of synchrotron emitting and
Faraday rotating layers along the line-of-sight, the relation between
$RM$ and $\mathcal{R}$ becomes more complicated. Sokoloff et al.
(\cite{sok98}) have calculated the result for various geometries of
synchrotron emitting and Faraday rotating layers. When the 
synchrotron emissivity and Faraday rotation are both constant per unit
line-of-sight (a so-called `Burn slab', Burn \cite{burn66}), 
$RM=0.5\mathcal{R_{\mathrm{max}}}$, where $\mathcal{R_{\mathrm{max}}}$
is the Faraday depth of the far side of the slab. Sokoloff et
al. showed that this is true for any distribution that is
mirror-symmetric along the line of sight.

We present a first discussion of the use of WSRT observations at
$\lambda \sim$ 0.86 m, of the linearly polarized component of the
diffuse Galactic synchrotron emission towards the Galactic
anti-centre. By using Faraday tomography we investigate the properties
of the magneto-ionic ISM in this direction, and in particular the
abundance of Faraday screens.

\section{The data}
The present dataset was obtained with the WSRT, a 14-element E-W
interferometer of which 4 elements are moveable to improve (u,v)
coverage. We used the mosaicking technique to map an area of about
7$\times$7 degrees$^2$ with 49 pointings. The distance between
pointings was chosen such that instrumental polarization is
suppressed to less than 1\% (Wieringa et al. \cite{wieringa93}). In
our analysis we leave out the edge of the mosaic where instrumental
polarization effects cannot be suppressed by mosaicking. We also
exclude lines-of-sight for which instrumental polarization of off-axis
sources is an important factor.  The central coordinates of this field
(in the constellation Gemini) are $\alpha = 7^h 18^m$ and $\delta =
36\degr 24\arcmin$ (J2000.0), which is $l \approx 181^{\circ}$ and $b
\approx 20^{\circ}$ in Galactic coordinates. The observations cover
the frequency range between 315 and 385 MHz, with 213 usable
independent spectral channels of about 0.4 MHz each.
The field was observed for 6 nights (@ 12 hrs each) in December 2002
and January 2003. This yielded visibilities at baselines from 36 to
2760 meters, with an increment of 12 meters. 
The results we discuss here were obtained with
a Gaussian taper that decreases the resolution to
2.2\arcmin\ $\times$\ 3.7\arcmin\ (FWHM at 350 MHz).

The data were reduced using the $\mathtt{NEWSTAR}$ data reduction package.
Dipole gains and phases and leakage corrections were determined using
the unpolarized calibrators 3C48, 3C147 and 3C295. Due to an a-priori
unknown phase offset between the horizontal and vertical dipoles,
signal can leak from Stokes U into Stokes V. We corrected for this by
rotating the polarization vector in the Stokes (U,V) plane back to the U
axis, assuming that there is no signal in V. The polarized
calibrator sources 3C345 and DA240 defined the sense of
derotation (i.e. to the positive or negative U-axis).  Special care
was taken to avoid automatic flagging of real signal on the shortest
baselines.
From 2
lines-of-sight with a strong polarized signal we estimate
that the amounts of ionospheric Faraday rotation in the 6 nights are identical
to within $\sim 10\degr$ 
so we did not correct for that.


Structure on large angular scales in Stokes Q and U will have been filtered out because we have no information on baselines below 36 m. We discuss the influence of this on our results in Sect. \ref{results}.


\section{Analysis}
\subsection{Methods}

\begin{figure}[t]

\resizebox{\hsize}{!}{\includegraphics[width=8.5cm]{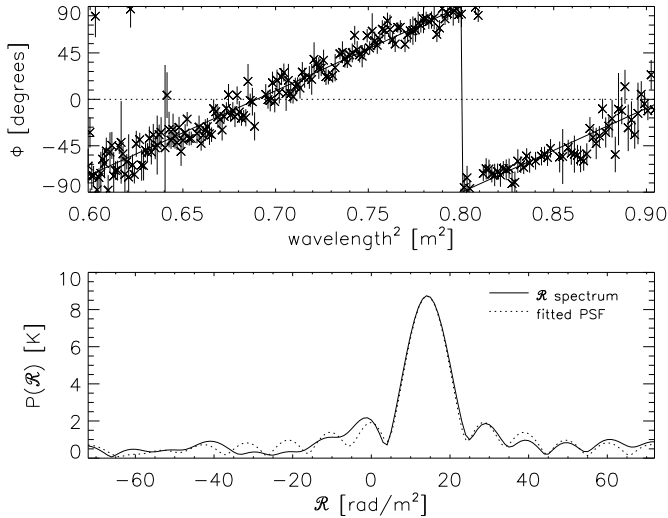}}

\caption{$\Phi(\lambda^2)$ distribution (top panel) and $\mathcal{R}$ spectrum (bottom panel) for a line-of-sight with a main peak that is similar to a Faraday screen (low $\Delta$). The $\chi^2_{\mathrm{red}}$ of the $RM$ fit is 1.02 (determined by using the $RM$ with the maximum $P(RM)$ - see the text).
}
\label{low_chi2red}
\end{figure} 

The wide coverage in $\lambda^2$ space (from 0.6m$^2\lesssim
\lambda^2\lesssim 0.9 $m$^2$), combined with the relatively small
channelwidth, allows us to do Faraday tomography, also known as
Rotation Measure Synthesis (see e.g. Brentjens \& De Bruyn
\cite{brentjensbruyn05}), which probes the distribution of Faraday
rotating and synchrotron emitting regions along the line-of-sight. In
Faraday tomography the polarization vectors of the individual channels
are `coherently added' by derotating the vectors using an assumed
Faraday depth $\mathcal{R}$: $\vec{P}(\mathcal{R})=\int
P(\lambda^2)\ \mathrm{e}^{2\mathrm{i}\Phi(\lambda^2)}
\mathrm{e}^{-2\mathrm{i}\mathcal{R}\lambda^2}\mbox{d}\lambda^2$, where
$P(\lambda)$ and $\Phi(\lambda)$ are the polarized intensity and
polarization angle of an individual channel at wavelength
$\lambda$. $P(\mathcal{R}) = |\vec{P}(\mathcal{R})|$ is the intensity
of the polarized emission at Faraday depth $\mathcal{R}$, and
a Faraday depth spectrum (or $\mathcal{R}$ spectrum) can be
constructed by calculating $P(\mathcal{R})$ and
$arg(\vec{P}(\mathcal{R}))$ for many values of $\mathcal{R}$.  We
calculated $\mathcal{R}$ spectra between -72 rad/m$^2$ and +72
rad/m$^2$ in steps of 1 rad/m$^2$. Previous surveys indicate that
the $\mathcal{R}$ of the diffuse emission in this volume of the Galaxy
lie well within this range (see e.g. Spoelstra \cite{spoelstra84}).
The Point Spread Function (PSF) of the $P(\mathcal{R})$-determination,
which is the Fourier Transform of the $\lambda^2$-sampling of the
data, has for our data a FWHP of about 12 rad/m$^2$.  From simulations
of the noise in the individual channels we estimate that 99\% of the
noise realisations will lie below the 0.7 K level in our $\mathcal{R}$
spectra.

We quantified the behaviour of the $\mathcal{R}$ spectrum in two ways.

First, we derived a measure of
the symmetry of the $\mathcal{R}$ spectrum as a whole, 
by calculating the reduced $\chi^2$ ($\chi^2_{\mathrm{red}}$) of a linear
fit to the $\Phi(\lambda^2)$ data.  A Faraday screen or a symmetric
distribution of Faraday rotating and synchrotron emitting regions
along the line-of-sight (for example a Burn slab) will show a linear
$\Phi(\lambda^2)$ dependence. The $\chi^2_\mathrm{red}$ of a linear
fit to the $\Phi(\lambda^2)$ distribution can therefore separate
symmetric and more complex distributions along the line-of-sight.

Fitting periodic data like polarization angles requires that the $n
180\degr$ periodicity of $\Phi(\lambda)$ is properly taken into
account. In Schnitzeler et al. (\cite{dominic07}) we introduced a
method that finds the lowest-$\chi^2_\mathrm{red}$ straight-line
fit to polarization angle data by going through all the possible
180\degr\ wraps of polarization angles that are allowed by the data. 
However, the number of configurations of wraps of the individual
datapoints increases strongly with the number of datapoints. For the
present dataset, with of order 200 $\Phi(\lambda)$-values for each
line of sight, application of that method is thus not practical.

However, by writing $\Phi(\lambda^2)$ as the complex number $\mathrm{e}^{2\mathrm{i}\Phi(\lambda^2)}$, 
the $RM$ spectrum can be calculated: $P(RM)=|\vec{P}(RM)|=|\int
\mathrm{e}^{2\mathrm{i}\Phi(\lambda^2)}\mathrm{e}^{-2\mathrm{i}RM\lambda^2}\mbox{d}\lambda^2|$.
The power $P(RM)$
shows which $RM$ `frequencies' create the observed $\Phi(\lambda^2)$
dependence. If $\Phi$ depends linearly on $\lambda^2$, the $RM$
with the maximum $P(RM)$ will be the best fitting slope for the
data. 
If the $RM$ spectrum is more complicated, the $RM$ with the maximum $P(RM)$ will not give the best linear fit to the $\Phi(\lambda^2)$, and the $\chi^2_{\mathrm{red}}$ of this fit will be higher than that of the best linear fit to the $\Phi(\lambda^2)$. 
Even the best linear fit to the $\Phi(\lambda^2)$ will then have a 
large $\chi^2_{\mathrm{red}}$.


\begin{figure}[t]

\resizebox{\hsize}{!}{\includegraphics[width=8.5cm]{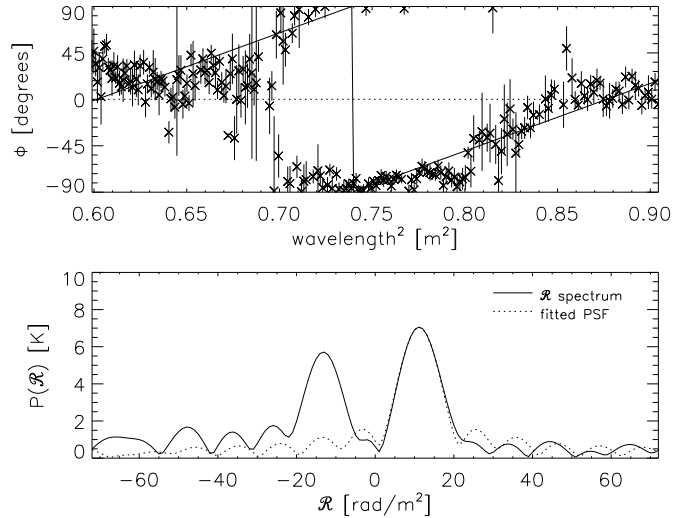}}

\caption{Identical to Fig. \ref{low_chi2red}. However, here the $\chi^2_{\mathrm{red}}$ of the $RM$ fit is 10.58, which is due to a complicated $\mathcal{R}$ spectrum.
}
\label{high_chi2red}
\end{figure} 



The second criterion quantifies the deviation of the main peak in the $\mathcal{R}$ spectrum from a Faraday screen, 
in the presence of noise.
We define $\Delta$ as the root-mean-square vertical separation in the $\mathcal{R}$ spectrum between a peak in the $\mathcal{R}$ spectrum and the PSF that is scaled to the same height as the peak.
$\Delta$ is calculated over the PSF out to the point where the PSF goes through its first minimum, at a distance of 10 rad/m$^2$ from the centre of the peak.
A Faraday screen will show up as an unresolved peak in the $\mathcal{R}$ spectrum, which means that its $\Delta$ will be lower than that of a peak in the $\mathcal{R}$ spectrum that is too broad to be fitted by a PSF. By comparing the $\Delta$ we find for the main peak 
in each line-of-sight to the distribution of $\Delta$ of a Faraday screen + noise, we can quantify how improbable it is that the main peak is due to a Faraday screen.


\begin{figure}[t]
\resizebox{\hsize}{!}{\includegraphics[width=8.5cm]{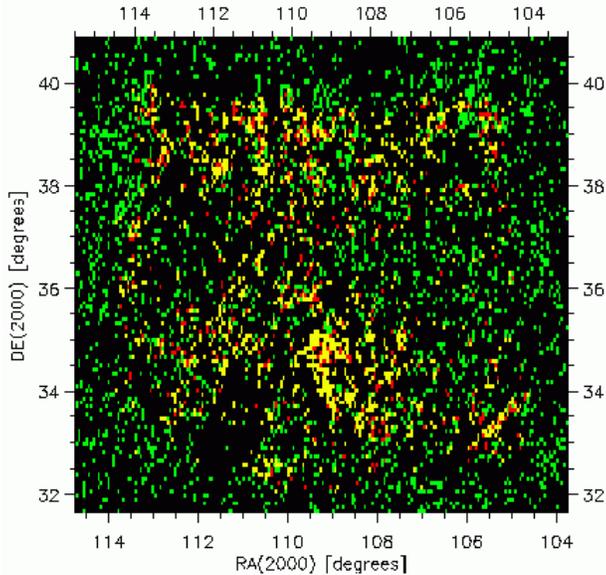}}

\caption{Distribution of independent lines-of-sight with a main peak that satisfies the $\Delta$ criterion (green pixels), or with $\chi^2_{\mathrm{red}} < 2$ and  max($P(\mathcal{R})$) $>$ 2 K (yellow pixels), or that satisfy all three criteria (red pixels). Lines-of-sight that did not pass these criteria are shown as black pixels. The pixel size is 2.2\arcmin$\times$2.2\arcmin$\csc(\delta)$. The outer edges of the mosaic show an increase in $\chi^2_{\mathrm{red}}$, which explains the absence of yellow pixels.
}
\label{fscreens}
\end{figure}


In Figs. \ref{low_chi2red} and \ref{high_chi2red} we show the $\Phi(\lambda^2)$ distributions and $\mathcal{R}$ spectra for two lines-of-sight that have a main peak that is similar to a Faraday screen (i.e. low $\Delta$). One of the lines-of-sight has a simple $\mathcal{R}$ spectrum and low $\chi^2_{\mathrm{red}}$, the other has a complicated $\mathcal{R}$ spectrum and high $\chi^2_{\mathrm{red}}$. Fig. \ref{high_chi2red} shows that low $\Delta$ are also found for lines-of-sight with higher $\chi^2_{\mathrm{red}}$. In this case the high $\chi^2_{\mathrm{red}}$ is due to the presence of multiple components in the $\mathcal{R}$ spectrum.

\subsection{Results}\label{results}
\begin{figure}[t]
\resizebox{\hsize}{!}{\includegraphics[width=12cm]{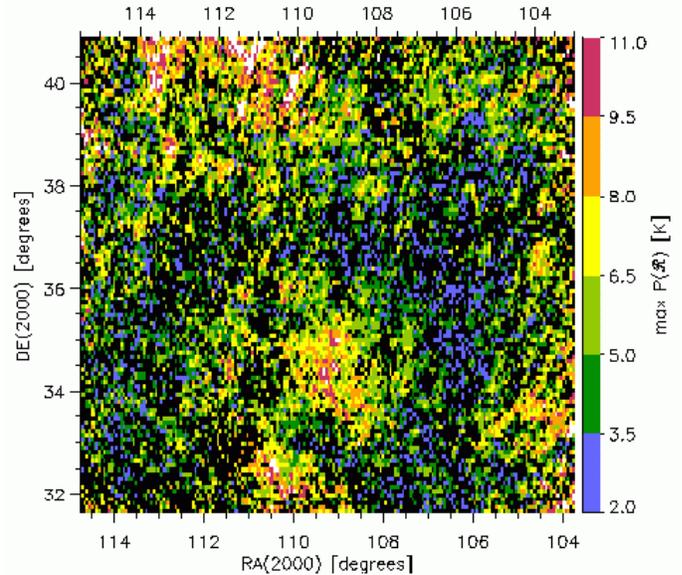}}

\caption{$P(\mathcal{R})$ of the highest peak in the $\mathcal{R}$ spectrum for lines-of-sight where the secondary peak in the $\mathcal{R}$ spectrum has at most half the strength of the main peak. Lines-of-sight that do not pass the selection criteria are shown in black, together with 455 lines-of-sight that have $P(\mathcal{R}) < $ 2 K.
}
\label{p_cr_derot_lm}
\end{figure}

\begin{figure}[t]
\resizebox{\hsize}{!}{\includegraphics[width=12cm]{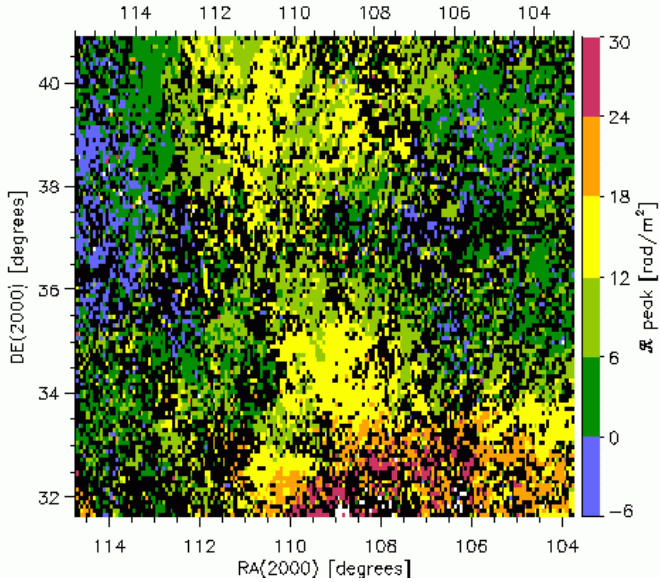}}

\caption{ 
$\mathcal{R}$ corresponding to the $P(\mathcal{R})$ plotted in Fig. \ref{p_cr_derot_lm}. The lines-of-sight plotted in these figures are selected by using the same criteria.
Lines-of-sight indicated in white have $\mathcal{R} >$ 30 rad/m$^2$. Lines-of-sight indicated in black either did not pass the selection criteria or have $\mathcal{R} <$ -6 rad/m$^2$ (652 and 9734 lines-of-sight resp.).
}
\label{cr_lm}
\end{figure}

From a total of 22.800 independent lines-of-sight in our dataset, 1757
lines-of-sight have a $\chi^2_{\mathrm{red}} < 2$ and a maximum in the
$\mathcal{R}$ spectrum $>$ 2 K, about 2.5 times the level below which
99\% of the noise $P(\mathcal{R})$ lie. In Fig. \ref{fscreens} we plot
the sky distribution of these 1757 lines-of-sight as yellow pixels. We
indicate the lines-of-sight that furthermore have a main peak whose
$\Delta$ is less than 99$\%$ of the $\Delta$ that we found in our
simulations of a Faraday screen + noise as red pixels. The pixels in
Figs. \ref{fscreens} to \ref{cr_lm} indicate independent
lines-of-sight. These figures cover Galactic longitudes from
($l,b$)=(178\degr,26\degr) in the top left corner to
($l,b$)=(185\degr,15\degr) in the bottom right corner. The Galactic
plane has a position angle (north through east) of about 21\degr\
relative to the vertical axis in these plots.

In Figs. \ref{p_cr_derot_lm} and \ref{cr_lm} we plot the distribution on the sky of $P(\mathcal{R})$ and $\mathcal{R}$ of the main peak in the $\mathcal{R}$ spectrum. Ionosphere models indicate that it contributes only 0.6 rad/m$^2$  to the $\mathcal{R}$ shown in Fig. \ref{cr_lm} (Johnston-Hollitt, priv. comm). For these maps we selected lines-of-sight where the $\mathcal{R}$ spectrum is dominated by a main peak, by requiring that this peak should be at least twice as high as the second brightest peak. Lines-of-sight selected in this way show a tight correlation between $\mathcal{R}$ and $RM$, with a 1$\sigma$ scatter of about 0.2 rad/m$^2$. 

Structure on large angular scales in Stokes Q and U (and therefore in polarized intensity $P$ and polarization angle $\Phi$) will not be picked up by an interferometer. However, if large-scale structure were produced at the same $\mathcal{R}$ as the emission visible in Fig. \ref{p_cr_derot_lm}, the large $\mathcal{R}$ gradients per pointing in Fig. \ref{cr_lm} create structure in $\Phi$ on small enough angular scales that can be picked up by the WSRT. Large-scale structure would then become `visible' by passing through the Faraday rotating foreground with these $\mathcal{R}$ gradients, and Figs. \ref{p_cr_derot_lm} and \ref{cr_lm} therefore form more or less `complete' maps in that they contain most of the emission at those Faraday depths.

\subsection{Faraday screens}\label{sec_fscreens}
The main peak in 3086 lines-of-sight (14$\%$ of the total number of independent lines-of-sight)\hspace{0.2mm} has a $\Delta$ that is less than 99$\%$ of the $\Delta$ we found in our simulations of a Faraday screen + noise, making these main peaks indistinguishable from Faraday screens. These lines-of-sight are indicated as green pixels in Fig. \ref{fscreens}. These peaks also closely follow the $\mathcal{R} = RM$ relation expected for Faraday screens, with a 1$\sigma$ scatter of about 0.2 rad/m$^2$. 
The actual fraction of lines-of-sight with Faraday screen components in their $\mathcal{R}$ spectrum can be higher than 14$\%$ since we only tested the main peak of the $\mathcal{R}$ spectrum with our $\Delta$ criterion.



Radio frequency interference and/or an underestimation of the noise levels will increase the values of $\Delta$ expected for a Faraday screen + noise. The lines-of-sight that already pass this criterion with our conservative choice for calculating $\Delta$ will then still pass the $\Delta$ criterion; there could however be more lines-of-sight that would qualify as Faraday screens.



\section{Discussion}

We first estimate the implied value of the large-scale magnetic field,
for a representative value of $\mathcal{R} =+10$ rad/m$^2$
(Fig. \ref{p_cr_derot_lm}).  If the Faraday rotating region has a
depth $L$ and consists of clumps of electrons with a constant density
$n_e$ that occupy a fraction $f$ of $L$, with $n_e$=0.08 cm$^{-3}$ and
$f$=0.4 (Reynolds \cite{reynolds91}), $L$ follows from observations of
the emission measure, $EM$. The $EM$ from the Wisconsin H$\alpha$
mapper (WHAM; Haffner et al. \cite{haffner03}) are on average 2
Rayleigh in this region, or 4 cm$^{-6}$pc for a 10$^4$ K gas, which
indicates a line-of-sight of 1.5 kpc length.
For these lines-of-sight extinction of the H$\alpha$ line plays only a
minor role.  The $DM$=50 cm$^{-3}$pc implied by this model agrees well
with the $DM$=54 cm$^{-3}$pc calculated from the NE2001 model by
Cordes \& Lazio (\cite{cordeslazio03}) for a line-of-sight in the same
direction and of the same length.

For this model the integral for $\mathcal{R}$ is also easy to solve. By using the same $n_e, f$ and $L$, the 
large-scale magnetic field component along the line-of-sight, $B_{\mathrm{reg},\|}$, must be about 0.2 $\mu$G to explain the average $\mathcal{R}$=10 rad/m$^2$ present in our data. If part of the H$\alpha$ emission is coming from beyond the Faraday rotating region, we would overestimate the depth $L$ of the Faraday rotating layer, and the $B_{\mathrm{reg},\|}$ we derive would be too low. A reasonable upper limit for $B_{\mathrm{reg},\|}$ can be found by adopting a pitch angle of 8\degr\ and a total strength of 4 $\mu$G for the large-scale field (both values from Beck \cite{beck07}), which gives $B_{\mathrm{reg},\|,\mathrm{max}} \approx 0.6$ $\mu$G.
Note that the change of sign of $\mathcal{R}$ in Fig. \ref{cr_lm} indicates that the observed features are also partly due to variations in the magnetic field geometry and not only to variations in the electron density.

From the Haslam et al. (\cite{haslam82}) 408 MHz data we estimate that
the Galactic foreground in this direction has a brightness temperature
of 36 K at 350 MHz, assuming a $-2.7$ spectral index for the
synchrotron brightness temperature. 
For a completely uniform magnetic field and without internal Faraday rotation the peak in the $\mathcal{R}$ spectrum would have a polarized brightness temperature, $T_{\mathrm{b,pol}}$, of 70\% of this, or 25 K, but
the $T_{\mathrm{b,pol}}$ of the peak in the $\mathcal{R}$ spectrum is
on average only 5 K. 
A similar discrepancy between the expected and
measured
$T_{\mathrm{b,pol}}$ was noted for a field observed with the WSRT at
($l,b$)=(110\degr,71\degr) (De Bruyn et al. \cite{debruyn06}).

If the $\mathcal{R}$ distribution has a finite width, the 25 K is diluted over the entire width, and the peak of the distribution will have lower $P(\mathcal{R})$. Any further reduction of $P(\mathcal{R})$ requires a turbulent magnetic field in the emitting region.
For lines of sight with a single, essentially unresolved peak in the
$\mathcal{R}$ spectrum (i.e. with a low value of $\Delta$) it is not
trivial to separate the two effects.  We simulated a layer with both
uniform and Gaussian distributions of coexisting Faraday rotation and
synchrotron emission. From these simulations we conclude that it is
not possible to reduce the $T_{\mathrm{b,pol}}$ of the maximum in the
$P(\mathcal{R})$ spectrum below 20 K, and be consistent with the
observed low value of $\Delta$.  Therefore a turbulent magnetic field
component is required to explain the observed $T_{\mathrm{b,pol}}$.

Another possible picture for these lines of sight with small $\Delta$
could be a combination of an $\mathcal{R}$-extended structure and an
emission region with a turbulent magnetic field. However, since the
$\mathcal{R}$-extended structure has to be narrow to pass our $\Delta$
criterion, the depth of this structure will not be enough to explain
the 10 rad/m$^2$ average Faraday depth in Fig. \ref{p_cr_derot_lm}. A
Faraday screen is then required to produce the bulk of this Faraday
depth.

An extreme example of this would be an emission region with a
turbulent field, observed through a Faraday screen, which would be the
simplest and most straightforward explanation. In that case we can
also estimate the relative strengths of the turbulent magnetic field
and the magnetic field component perpendicular to the line-of-sight,
$B_{\mathrm{turb}}$ and $B_{\mathrm{reg,\perp}}$ resp. If the
synchrotron emitting region contains both $B_{\mathrm{reg,\perp}}$ and
$B_{\mathrm{turb}}$, the expected polarization fraction is reduced by
a factor of
$B_{\mathrm{reg,\perp}}^2/(B_{\mathrm{reg,\perp}}^2+B_{\mathrm{turb}}^2)$
(Burn \cite{burn66}). To reduce the expected 25 K polarized brightness
temperature to 5 K then requires
$B_{\mathrm{turb}}/B_{\mathrm{reg,\perp}} \approx 2$.
This ratio is an upper limit if there are other components in the $\mathcal{R}$ spectrum that contribute to the predicted $T_{\mathrm{b,pol}}$ = 25 K.

The region around ($\alpha,\delta$)=109\degr,35\degr shows some conspicuous features. It is bright in polarized intensity, shows a low $\chi^2_{\mathrm{red}}$, and also contains a number of lines-of-sight that could be Faraday screens. The band between 36\degr\ $\lesssim \delta \lesssim$ 38\degr\ contains many lines-of-sight with high $\chi^2_{\mathrm{red}}$ and broad main peaks and/or significant secondary peaks in the $\mathcal{R}$ spectrum. We will discuss these regions in a future article. 


\section{Conclusions}
We used high spectral resolution radio polarization data to study the Galactic ISM, and we demonstrated a number of tools that can be used for this purpose. In the present dataset we identified a significant number of lines-of-sight that show an unresolved peak in their Faraday depth spectrum, similar to a Faraday screen. We also studied the spatial behaviour of the principal component in the Faraday depth spectrum, and for the lines-of-sight that only have one unresolved component in their $\mathcal{R}$ spectrum, we showed that a model of a Faraday rotating region in front of a synchrotron emitting region that contains a turbulent magnetic field component can explain the data.

\begin{acknowledgements}
We would like to thank Dan Stinebring for his comments on an earlier version of the manuscript. DS would also like to thank Russ Taylor and Jeroen Stil for pointing out a flaw in an earlier version of the `missing large-scale structure' argument.
The Westerbork Synthesis Radio Telescope is operated by the
Netherlands Foundation for Research in Astronomy (NFRA) with financial
support from the Netherlands Organization for Scientific Research
(NWO). The Wisconsin H-Alpha Mapper is funded by the National Science
Foundation.
\end{acknowledgements}

\end{document}